\documentclass[twocolumn,showpacs,preprintnumbers,amsmath,amssymb,subeqn,superscriptaddress]{revtex4}

\usepackage{graphicx}
\usepackage{dcolumn}
\usepackage{bm}

\begin{document} 

\selectfont

\title{Competitive cluster growth in complex networks}

\author{Andr\'e A. \surname{Moreira}}
\affiliation{Departamento de F\'{\i}sica, Universidade Federal do
             Cear\'a,\\ 60451-970 Fortaleza, Brazil}
\email{auto@fisica.ufc.br}
\author{Dem\'etrius R. \surname{Paula}}
\affiliation{Departamento de F\'{\i}sica, Universidade Federal do
             Cear\'a,\\ 60451-970 Fortaleza, Brazil}
\author{Raimundo N. \surname{Costa Filho}}
\affiliation{Departamento de F\'{\i}sica, Universidade Federal do
             Cear\'a,\\ 60451-970 Fortaleza, Brazil}
%\homepage[URL: ]{http://www.fisica.ufc.br/demas}
\author{Jos\'e S. \surname{Andrade}, Jr.}
\affiliation{Departamento de F\'{\i}sica, Universidade Federal do
             Cear\'a,\\ 60451-970 Fortaleza, Brazil}

\date{\today}
 
\begin{abstract}
Understanding the process by which the individuals of a society make
up their minds and reach opinions about different issues can be of
fundamental importance. In this work we propose an idealized model
for competitive cluster growth in complex networks. Each cluster can
be thought as a fraction of a community that shares some common
opinion. Our results show that the cluster size distribution depends
on the particular choice for the topology of the network of contacts
among the agents. As an application, we show that the cluster size
distributions obtained when the growth process is performed on
hierarchical networks, e.g., the Apollonian network, have a scaling
form similar to what has been observed for the distribution of number
of votes in an electoral process. We suggest that this similarity is
due to the fact that social networks involved in the electoral process
may also posses an underlining hierarchical structure.
\end{abstract}

%%%%PACS e Keywords
 %Structures and organization in complex systems


\maketitle

%%%%%%%%%%%%%%%%%%%%% INTRODUCTION
%{\bf Introduction}
%\section{Introdu\c{c}\~ao}

The sedimentation of new trends and ideas in large social communities
can have a profound impact in the life of individuals. An instance
where the dynamics of opinion formation may be o major importance is
the democratic elections of representatives. In addition, in the
electoral process every agent is called to give her/his opinion in an
anonymous way and the statistical results are easily
accessible~\cite{costa99}. This makes elections ideal systems to
researchers interested in studying the process of opinion
formation. Since, typically, an individual is more likely to listen to
someone they have a personal contact, the process may also be driven
by a mouth-to-mouth interaction besides from massive political
campaigns based on media programs. In this way, the spreading of an
opinion follows a pattern similar to the spreading of an epidemic
process~\cite{gladwell00}. One should also expect that the particular
structure of the network of contacts among the agents of a society may
have an impact in the way the opinions propagate.

The intricate structure of interactions of many natural and social
systems has been object of intense research in the new area of complex
networks. Most of the effort in this area has been directed to find
the topological properties of real world
networks~\cite{watts98,barabasi99,albert99,amaral00,camacho02} and
understand the effects that these properties cast on dynamical
processes taking place on these complex
networks~\cite{albert00,watts02,castellano03,gonzalez04}. For
instance, the small-world characteristic~\cite{watts98}, where each
node of the network is only a few connections apart from any other,
permits a quick spreading of information through the network, being
fundamental in processes of global coordination~\cite{moreira04} and
feedback regulation~\cite{amaral04}.

Another property commonly studied in complex networks is the degree
distribution $P(k)$, that gives the probability with which an
arbitrary node is connected to exactly $k$ other nodes. One relevant
characteristic often observed in complex networks is a scale-free
degree distribution~\cite{barabasi99}, namely, a distribution which
follows a power-law, $P(k) \sim k^{-\gamma}$, with an exponent
typically in the range $2<\gamma <3$. Such broad degree distribution
has a dramatic effect in many dynamical processes. In the spreading of
infectious diseases, for example, it has been shown that when the
infection is mediated by a scale-free network, any infection rate
above zero results in a positive fraction of infected
individuals~\cite{pastor-satorras01}. It was recently suggested that
another universal characteristic of real-world networks is a structure
of communities, where smaller communities in the network are joined to
larger communities by highly connected nodes that play the role of
local hubs~\cite{ravasz02}. This structure may be related to the
self-similar characteristic observed in some
complex-networks~\cite{song05}.

\begin{figure}[t]
\vspace{1cm}
\centering
\includegraphics[width=7.9cm]{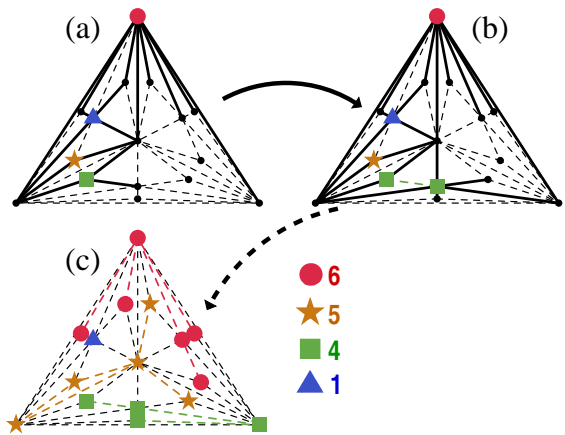}
\caption{Pictorial description of our
model for a competitive cluster growth process. 
To make a clear picture of our model, we used as a substrate network
for the growth a third generation Apollonian
network~\protect{\cite{andrade05}}. The matching graph form of this
network makes it particularly suitable for a plane representation.
In the first step of the growth model (a), a few nodes (the large
circle, square, triangle, and star) are chosen to be the seeds of the
growing clusters. Each of the seeds is the first node of a different
cluster. All the remaining nodes (small dots) do not assigned to any
of the clusters in the beginning, this nodes will be the ones
accessible to the growth.
The dashed lines link either a pair of nodes that already belonging to
one of the clusters or a pair of nodes that are not in any cluster
yet. At this time-step these lines do not participate in the growth
process.
The thick lines link one of the seeds to an accessible node,
and any of the thick lines can be chosen with the same
probability to channel the growth of a cluster. In this way,
the growth rate of a particular cluster is proportional to
its perimeter, that is, the number of connections from one
of the nodes already incorporated into the cluster to an
accessible node.
In the second step (b) a new node is incorporated the cluster of
squares and new thick lines are added to the perimeter of this
cluster.
The process continues until every one of the nodes have been
incorporated to one of the clusters (c). At this point the size of all
the clusters is computed.
This mechanism of cluster growth is a greedy process. The larger the
perimeter of a cluster faster it will grow incorporating more nodes
and more connections.}
\label{f.m}
\end{figure}

In this letter we investigate a dynamic process of competitive cluster
growth in complex networks.
In this process many alternative and self-excluding states are
accessible to the nodes of the network. We say that nodes in the same
state belong to a cluster, and each of these clusters competes with
the others to reach a larger part of the network.
This idealized mechanism can be thought as a model for a variety of
different processes that take place in real networks. For example, one
can think of each cluster as the part of a population that has been
infected with a certain strain of a virus. Alternatively, the clusters
may represent alternative opinions in a social group.
We study the distribution of the fraction of the network occupied by
an arbitrary cluster. Our results show that the network topology has
great influence over the behavior of the cluster size distributions.

\begin{figure}[t]
\vspace{1cm}
\centering
\includegraphics[width=7.9cm]{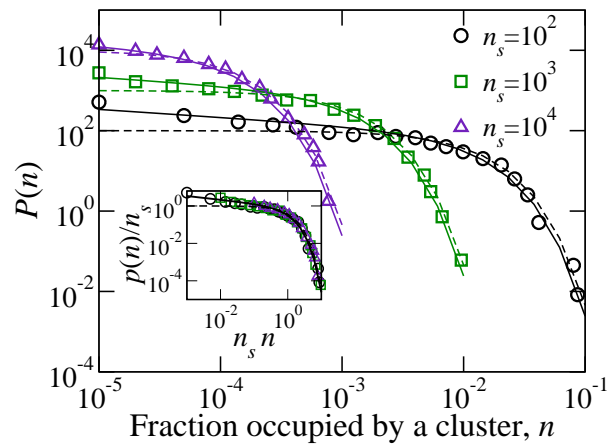}
\caption{Normalized cluster size distributions obtained 
for the competitive growth model under a random ER network topology.
These distributions were obtained for networks of $10^5$ nodes. We
performed a larger number of runs for the cases with less competing
clusters in order to have $10^5$ cluster samples for each case.
The distributions obtained under this topology follow approximately an
exponential distribution with a characteristic cluster size depending
on the number $n_s$ of competing clusters.
The dashed lines are the functions $n_s e^{-n_s n}$. The continuous
lines are a fit for the function
$P(n)=C n_s^{1-\gamma} n^{-\gamma} \exp(-n_s n)$,
with the parameters $C=0.85 \pm 0.03$, and $\gamma=0.2 \pm 0.05$
obtained for the collapsed data shown in the inset.
This figure shows that for the random networks the form of the
distribution is completely determined by the density of nodes chosen
to be seeds for the growth process.}
\label{f.r}
\end{figure}

Our model for competitive cluster growth is described as follows: the
process is mediated by a substrate network with $N$ nodes; in the
first moment, a number $n_{s}$ of nodes is chosen at random to be the
seeds of the spreading process, with the density of seeds being
$n_s/N$. Each seed will be the first node of a cluster.
Then, the clusters grow by incorporating nodes that are neighbors of
these seeds and have not yet been assigned to any other cluster.
Once a node is incorporated to a cluster it will stay in this cluster
until the end of the process, and only the nodes not belonging to any
of the existing clusters are accessible to the growth process. We
will refer to these as accessible nodes.
The growth process takes place in discrete steps.
At each step, we randomly select a pair of connected nodes,
one belonging to a cluster and the other that is accessible to
growth. 
The accessible node is subsequently incorporated to the same cluster
as its neighbor.
In this way, the average rate of growth of a cluster is proportional
to the number of accessible nodes in its perimeter. The process
continues until all the nodes have been incorporated to one of the
competing clusters and no accessible node is left on the
network. Figure~\ref{f.m} presents a pictorial description of our
growth model.
We suggest that this mechanism can be representative of some sort of
greedy process where large clusters, with many nodes, will have more
connections to accessible nodes and therefore tend to grow faster and
increase even more their perimeters.
In this way, our model resembles the preferential attachment model for
network growth~\cite{barabasi99}. However, the competitive growth has
significant differences from that mode, in the sense that all seeds
are present in the beginning of the process and also the fact that
some clusters will start to grow before others, resulting in a variety
of cluster sizes.

\begin{figure}[t]
\vspace{1cm}
\centering
\includegraphics[width=7.9cm]{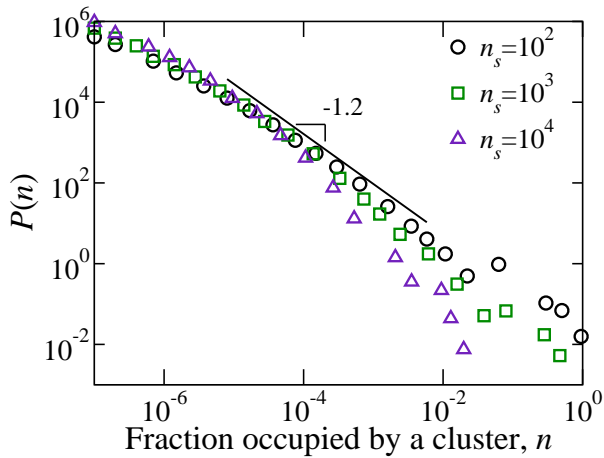}
\caption{Normalized cluster size distributions obtained 
when the growth process is performed in networks built with the
preferential attachment schema. We used networks with $10^7$ nodes and
obtained $10^5$ cluster samples.
Different from the the case of random networks, here we can not
produce a data collapse for the distributions. We note, however, that
for any value of $n_s$ we have a broad distribution with clusters of
all sizes ranging from just a few nodes to most of the network.
The continuous line is a power-law fit, $P(n) \sim n^{-\alpha}$, with
the exponent $\alpha=1.2$ obtained for the scaling region of the
distribution for $n_s=10^2$. The slope of the becomes more steep as
the number of competing clusters in the network grows.}
\label{f.b}
\end{figure}

The choice of a particular topology for the network of contacts should
affect the growth process. The simplest model for a network is
probably the random network model proposed by Erdos and Reny
(ER)~\cite{erdos}. In this model, any pair of nodes can be connected
with probability $\rho$, and the degree distribution follows
approximately a Poisson form with the average degree given by
$\bar{k}=\rho(N-1)$. In Fig.~\ref{f.r} we show the distribution of the
fractions of the network corresponding to each cluster $P(n)$ when the
spreading process takes place on ER networks. In this case the cluster
size distributions are power-laws with a small exponent bounded by an
exponential cutoff with a characteristic scale depending only on the
density of competing clusters in the network.

%This result can be understood when one notes that
%the average cluster size $N/n_s$ is the inverse of the
%density of seeds.

Next, we study the cluster growth process on a scale-free network.
To build the scale-free network we use the so-called preferential
attachment method~\cite{barabasi99}. Different from the random graph,
the degree distribution of the networks built with this model have a
power-law form, $p_k(k) \sim k^{-\gamma}$, with an exponent $\gamma=3$,
followed by an exponential cutoff at a maximum degree, $k_{max} \sim
N^{\frac{1}{\gamma-1}}$~\cite{krapvisky}. Note that, a cluster that
incorporates one of the most connected nodes in the beginning of the
growth process will increase its growth rate by a large factor. Thus,
the long tail of this distribution may have a dramatic effect on the
cluster growth process. Indeed, the cluster size distribution obtained
when the spreading process is done in the preferential attachment
network also shows a power-law tail, $p(n)\sim n^{-\alpha}$, with an
exponent $\alpha=1.2$ for the case with $n_s=10^2$~\footnote{Similar
results were obtained for the distributions of cluster size in the
transient state of the Snadj model~\protect\cite{bernardes02}.}, as
seen in Fig.~\ref{f.b}. Interestingly, the competing growth dynamics
magnifies the effect of the scale-free distribution producing
distributions of cluster sizes that have smaller exponents, and as a
consequence, a slower decay than that observed for the degree
distribution. The scattering of the data close to the limit $n \to 1$
is due to statistical fluctuations at this low frequency limit as well
as to finite size effects. The implication of these heavy tailed
distributions is that now the average cluster size is not a
characteristic scale for the process since one finds, with relatively
high frequency, clusters that are orders of magnitude larger than the
average.

\begin{figure}[t]
\vspace{1cm}
\centering
\includegraphics[width=7.9cm]{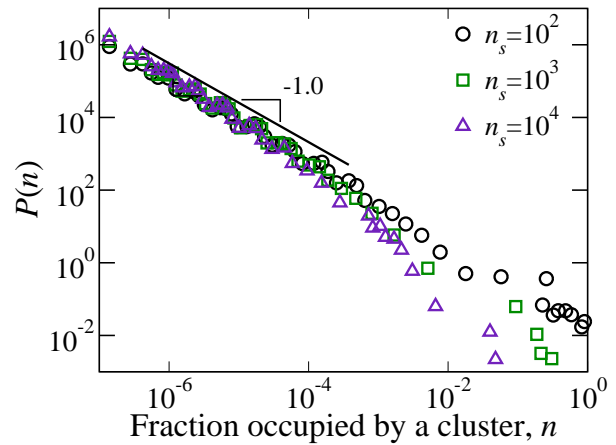}
\caption{Normalized cluster size distribution when our 
model is performed on Apollonian networks. The distributions were
obtained with the 15th generation of the Apollonian network what
corresponds to $N=7174456$. Each curve was sampled from an ensemble of
$10^5$ clusters. As in the case of the preferential attachment
networks here we observe a power-law decay with an exponent $\alpha
\approx -1$. The continuous line is a fit for the scaling region of
data with $n_s=100$ initial seeds. The periodical steps observed in
the shape of the distributions is remnant of the self similar
structure resulting from the hierarchical construction of the
Apollonian network.}
\label{f.a}
\end{figure}

Although the preferential attachment model produces a degree
distribution similar to what is found in several real networks, it
does not display the self-similarity and hierarchical structure also
observed in many of those networks. In the work of Ravasz et
al.~\cite{ravasz02}, it was suggested that the structure of a network
could be probed in a quantitative way by studying the cluster
coefficient of its nodes. The cluster coefficient of a node is defined
as the probability that two of its neighbors taken at random are
connected. A signature of the hierarchical structure would be a
cluster coefficient proportional to the inverse of the nodes degree
$c(k) \sim k^{-1}$~\cite{ravasz02}.
Both the preferential attachment and the random networks do not
present this property. A model that shows small-world behavior,
scale-free degree distribution as well as a hierarchical structure is
the recently proposed Apollonian network~\cite{andrade05}.

In order to test the effect of a hierarchical structure in the cluster
growth we implemented our model in the topology of the Apollonian
network. In Fig.~\ref{f.a} we show that the cluster size distributions
obtained for the clusters grown in this topology follow power-law
behavior with an exponent $\alpha=-1$.
Such behavior can be understood with a simple scaling argument.
Splitting the Apollonian network in the most connected hubs one finds
three smaller networks corresponding to Apollonian networks of a lower
generation. Each of these pieces could be split again producing nine
smaller networks with this hierarchical disassembly continuing down to
the level of single nodes.
After the growth process is performed in a large Apollonian network,
one could measure the cluster size distribution for each of these
generations. It should be expected that, as one goes to higher
generations, the distributions approach a limiting form.
Note also that only the clusters that reach the three hubs in the
corner of the networks are merged when one moves one generation up,
and only the size of these few clusters that reach these hubs change
from one generation to another. Thus, this process can change the form
of the distribution only in the limit of very large clusters, where the
frequency is of the order of the inverse of the network size.
This means that the distribution of cluster fractions should obey
approximately the following similarity relation,
\begin{equation}
P(n)=3P(3n).
\label{simi}
\end{equation}
A trivial solution for this similarity relation is the Dirac delta
$\delta(n)$. We expect the distribution to assume this form if, with
probability one, the growth process produces one giant cluster that
incorporates a large fraction, $n
\approx 1$, as $N$ grows.
Besides the delta function, any nontrivial function that satisfies the
similarity relation~(\ref{simi}) should be of the form $C/n$. Although
we have not proved that a giant cluster does not appear, our numerical
results indicate that, at least for the densities of seeds we have
investigated, this is unlikely.

Intriguingly, the scaling found for the distributions obtained with
the Apollonian network is similar to those found in the distribution
of number of votes per candidate in the elections in
Brazil~\cite{costa99}. This leads us to suggest that the universal
behavior observed in the electoral processes may by driven by an
underling hierarchical structure of the social networks describing the
interactions among voters.

In summary, we have introduced a model for competitive cluster growth
in complex networks. By means of numerical simulations we have shown
that the fraction of the network accessed by each cluster follows a
characteristic distribution that depends on the particular topology of
the network. For the case of a random graph we found an exponential
decay while for scale-free networks we found power-law decaying
distributions. In the particular case of a hierarchical scale-free
network, we observed a decay with a governing exponent $\alpha=-1$.
The distributions we obtained with our model resemble the
distributions of the fraction of votes per candidates observed in the
proportional elections in Brazil~\cite{costa99}. This leads us to
conjecture that an hierarchical structure may be present in the social
network governing the process by which voters reach their decisions.

We thank A. D. Ara\'ujo, D. M. Auto, and H. J. Herrmann for all the
discussions. The authors acknowledge the financial support of the
Brazilian agencies CNPq, CAPES, and FUNCAP.

\end{document}